\documentclass[aps,pra,10pt,amsmath,twocolumn,floatfix,eqsecnum,superscriptaddress,longbibliography]{revtex4-1}
\usepackage[dvipsnames]{xcolor}
\usepackage[colorlinks=true,bookmarks=false,linkcolor=NavyBlue,urlcolor=NavyBlue,citecolor=NavyBlue,breaklinks]{hyperref}
\usepackage{amsmath}
\usepackage{amsfonts}
\usepackage{amssymb}
\usepackage{tabularx}
\usepackage{graphicx}
\usepackage{times}
\usepackage{mathtools}
\usepackage{braket}
\usepackage{amsthm}

\newcommand{\avg}[1]{\langle#1\rangle}
\newcommand{\Avg}[1]{\left\langle#1\right\rangle}

\newcommand{\bk}[1]{\left(#1\right)}
\newcommand{\Bk}[1]{\left[#1\right]}
\newcommand{\BK}[1]{\left\{#1\right\}}

\newcommand{\norm}[1]{\lVert #1 \rVert}

\newcommand{\product}{\mathcal E}
\newcommand{\cp}{\mathcal F}
\newcommand{\gce}{\mathcal F_\sigma}

\DeclareMathOperator{\trace}{tr}
\DeclareMathOperator{\real}{Re}

\begin{document}

\title{Generalized conditional expectations for quantum retrodiction
  and smoothing}

\author{Mankei Tsang}
\email{mankei@nus.edu.sg}
\homepage{https://blog.nus.edu.sg/mankei/}
\affiliation{Department of Electrical and Computer Engineering,
  National University of Singapore, 4 Engineering Drive 3, Singapore
  117583}

\affiliation{Department of Physics, National University of Singapore,
  2 Science Drive 3, Singapore 117551}

\date{\today}

%\pacs{42.50.Wk, 03.65.Ta, 42.65.Yj}

\begin{abstract}
  The inference of a hidden variable's historical value, based on
  observations before and after the fact, is a controversial subject
  in quantum mechanics. Here I address the controversy by proposing a
  formalism that unifies and generalizes some of the previous
  proposals for the task, including the quantum
  minimum-mean-square-error estimators proposed by Ohki, the
  generalized conditional expectation proposed by Accardi and
  Cecchini, the quantum smoothing theory proposed by Tsang, the
  optimal observables for parameter estimation proposed by Personick,
  Belavkin, and Grishanin, and the weak values proposed by Aharonov,
  Albert, and Vaidman. The formalism is based on Ohki's suggestion of
  a distance between two observables in the Heisenberg picture, which
  remains well defined for incompatible observables and serves as a
  more general foundation for quantum inference than Belavkin's
  nondemolition principle.
\end{abstract}

\maketitle
\section{Introduction}
The inference of a hidden variable's historical value, based on
observations before and after the fact, is a fascinating yet
controversial subject in quantum mechanics; see, for example,
Refs.~\cite{watanabe,aav,*aharonov_rohrlich,*dressel14,gough20,gough20a,barnett21,*barnett00,*pegg02,chantasri21,leifer13,horsman17},
and references therein. This problem, called retrodiction or smoothing
in the engineering literature \cite{kalman,sarkka}, is well defined in
classical statistics and, indeed, a common and uncontroversial
endeavor in human activities, but its definition in quantum mechanics
is less settled. The core of the issue is the compatibility between
the inferred observable and the measured observable in the Heisenberg
picture. If the future value of an observable ahead of the measurement
is to be inferred, as in the prediction and filtering problem, the
compatibility holds, as per the seminal work of Belavkin
\cite{belavkin_qnd}.  The compatibility implies that the observables
can, in principle, be measured jointly, and the laws of classical
probability apply to their joint statistics. This so-called
nondemolition principle (NDP) underlies the quantum prediction and
filtering theory pioneered by Belavkin \cite{belavkin_qnd,bouten}. In
the retrodiction and smoothing problem, on the other hand, the
observables may not commute, and although many have proposed quantum
formulations of the task
\cite{aav,smooth,smooth_pra1,*smooth_pra2,*tsang14a,ohki15,*ohki18,chantasri21},
such as the weak values \cite{aav}, others have argued that the
inference does not make sense if it violates the NDP \cite{gough20}.

To clarify the murky state of affairs, here I propose a formalism that
unifies some of the existing attempts at the quantum retrodiction and
smoothing theory. The formalism is based on a principle of quantum
inference that supersedes the NDP. I start with Ohki's proposal of
quantum minimum-mean-square-error estimators based on certain inner
products \cite{ohki15}. The definition of an error, and an estimator
to minimize it, gives a concrete decision-theoretic meaning to the
inference problem, even if the involved observables do not commute.  I
then convert the Heisenberg-picture approach of Ohki to a formalism
based on open quantum system theory in the Schr\"odinger
picture. Remarkably, the optimal estimators then coincide with some of
the generalized conditional expectations (GCEs) that have been studied
in mathematical physics; see, for example,
Refs.~\cite{accardi82,petz86,petz88a,jencova10,hayashi}, and
references therein.  In the latter context, the concept has a long
history and has been found to be useful as an intermediate
mathematical tool, but there does not seem to be any attempt at
applying it to a physical quantum inference problem or relating it to
quantum retrodiction and smoothing.  The main goal of this paper is to
forge connections between the different areas. Many other prior works
on quantum inference
\cite{watanabe,barnett00,aav,smooth,smooth_pra1,gammelmark2013,guevara,personick71,belavkin73}
emerge as special cases of the formalism here.

% A concept analogous to the GCEs may be quasiprobability
% representations, which qualify as probability distributions only in
% special cases. Those special cases remain tremendously useful,
% however, in providing semiclassical models even with incompatible
% observables, and their failure to remain positive often serves as
% signatures of nonclassicality \cite{ferrie11}. As can be seen from the
% literature on weak values \cite{pusey14}, the GCEs may serve a similar
% purpose beyond their established mathematical role.

This paper is structured as follows. To set the stage,
Sec.~\ref{sec_review} reviews the established concepts of classical
conditional expectation and the NDP. Section~\ref{sec_gce} presents
the formalism of GCEs. Section~\ref{sec_hybrid} studies the
application of the GCEs to problems that obey the NDP, including the
classical and hybrid retrodiction and smoothing problems
\cite{watanabe,barnett00,smooth,smooth_pra1,gammelmark2013,ohki15,guevara}
and some quantum estimation problems \cite{personick71,belavkin73}.
Section~\ref{sec_nonclassical} presents some examples that may violate
the principle, namely, the weak values \cite{aav} and the application
of retrodiction and smoothing to linear Gaussian systems
\cite{smooth,smooth_pra1}.  Section~\ref{sec_conclu} is the
conclusion.

\section{\label{sec_review}Review of established concepts}
\subsection{\label{sec_ce}Classical conditional expectation}
Before discussing quantum generalizations of the conditional
expectation, I first review the concept in classical probability
theory \cite{parthasarathy05,billingsley}. Let $(\Omega,\Sigma,P)$ be
a probability space, where $\Omega$, the sample space, is the set of
all possible outcomes of an experiment, $\Sigma$ is a sigma-algebra
that consists of subsets of $\Omega$, and $P:\Sigma\to [0,1]$ is a
probability measure. If $A:\Omega \to \mathbb R$ is a Borel map that
models a real-valued random variable and $\Sigma_1$ is a
sub-sigma-algebra of $\Sigma$, then the expectation of $A$ conditioned
on $\Sigma_1$, denoted by $E(A|\Sigma_1):\Omega \to \mathbb R$, is a
$\Sigma_1$-measurable function defined by
\begin{align}
\int_{S_1} A(\omega) P(d\omega)
&= \int_{S_1} E(A|\Sigma_1)_\omega P(d\omega)
\quad
\forall S_1 \in \Sigma_1.
\end{align}
The conditional probability of an event $S \in \Sigma$ is then given
by $E(1_S|\Sigma_1)$, where $1_S$ is the indicator function
($1_S(\omega) = 1$ if $\omega \in S$ and $0$ otherwise).  To be more
concrete, suppose that $\Sigma_1$ is generated by an observed random
variable $Y:\Omega \to \mathcal Y$, in the sense of
\begin{align}
\Sigma_1 &= \BK{Y^{-1}(S): S \in \mathcal B_{\mathcal Y}},
\\
Y^{-1}(S) &\equiv \BK{\omega: Y(\omega) \in S},
\end{align}
where $\mathcal Y$ is a topological space and
$\mathcal B_{\mathcal Y}$ is the Borel sigma-algebra with respect to
$\mathcal Y$.  Then any $\Sigma_1$-measurable function is simply a
function that can be expressed in terms of another Borel map
$c:\mathcal Y \to \mathbb R$ as
\begin{align}
C(\omega) = c[Y(\omega)] \equiv (Y^* c)(\omega),
\end{align}
where $Y^*$ denotes the pullback. Henceforth, I denote the conditional
expectation as $E(A|Y)$ if $\Sigma_1$ is generated by $Y$.

If we restrict our attention to random variables with finite variance,
the conditional expectation can be defined in terms of Hilbert-space
theory \cite{parthasarathy05}. Define an inner product between two
real-valued random variables as
\begin{align}
\Avg{B,A}_P &\equiv \int B(\omega) A(\omega) P(d\omega),
\end{align}
the associated norm as
\begin{align}
\norm{A}_P &\equiv \sqrt{\Avg{A,A}_P},
\end{align}
and the distance between two random variables
as
\begin{align}
d_P(A,B) &\equiv \norm{A-B}_P.
\label{dP}
\end{align}
A Hilbert space $L_2(P)$ of random variables can then be constructed,
with each element corresponding to an equivalence class of random
variables with zero distance between them, while a subspace $L_2^Y(P)$
can be constructed from the $\Sigma_1$-measurable functions. $E(A|Y)$
can be defined as the $L_2^Y(P)$ element that satisfies
\begin{align}
\Avg{C,A}_P &= \Avg{C,E(A|Y)}_P
\quad
\forall C \in L_2^Y(P),
\label{inner_ce2}
\end{align}
which implies that $E(A|Y)$ is the projection of $A$ into
$L_2^Y(P)$. It follows from basic Hilbert-space theory that $E(A|Y)$ is
the $L_2^Y(P)$ element closest to $A$.

If $A$ is hidden and $B$ is the estimator given $Y$ in a Bayesian
inference problem, then the distance given by Eq.~(\ref{dP}) is the
root-mean-square error, and $E(A|Y)$ is the minimum-mean-square-error
estimator of $A$ given $Y$ \cite{parthasarathy05}.

Since $E(A|Y)_\omega$ is a $\Sigma_1$-measurable function, it can be
expressed in terms of a Borel map $\check a:\mathcal Y \to \mathbb R$
as
\begin{align}
E(A|Y)_\omega &= \check a[Y(\omega)].
\end{align}
Let $P_Y$ be the coarse-grained measure induced by $Y$, defined as
\begin{align}
P_Y(S) &\equiv P[Y^{-1}(S)],\quad S \in \mathcal B_{\mathcal Y}.
\end{align}
Then the right-hand side of Eq.~(\ref{inner_ce2}) can be expressed as
\begin{align}
\Avg{C,E(A|Y)}_P &= \int c(y) \check a(y) P_Y(dy)
= \Avg{c,\check a}_{P_Y},
\end{align}
and Eq.~(\ref{inner_ce2}) becomes
\begin{align}
\Avg{Y^* c,A}_P &= \Avg{c,\check a}_{P_Y}
\quad
\forall c \in L_2(P_Y).
\label{inner_ce3}
\end{align}
This equation for the conditional expectation turns out to be the most
convenient one for quantum generalizations.

To make a connection with more elementary probability theory, suppose
that $A$ can be expressed as a Borel map $a:\mathcal X \to \mathbb R$
of another random variable $X:\Omega \to \mathcal X$, viz.,
\begin{align}
A(\omega) &= a[X(\omega)],
\end{align}
and assume that the ranges $\mathcal X$ and $\mathcal Y$ of
$X(\omega)$ and $Y(\omega)$ are finite sets.  Let
\begin{align}
P_{XY}(x,y) &\equiv 
P\bk{\BK{\omega: X(\omega) = x \textrm{ and }Y(\omega) = y}}
\end{align}
be the joint probability distribution of $X$ and $Y$. Then
Eq.~(\ref{inner_ce3}) becomes
\begin{align}
\sum_{x,y} c(y) a(x) P_{XY}(x,y) = \sum_y c(y) \check a(y) P_Y(y)
\nonumber\\
\forall c \in L_2(P_Y).
\end{align}
The solution for $\check a(y)$ with $P_Y(y) > 0$ is
\begin{align}
\check a(y) &= \sum_x a(x) \frac{P_{XY}(x,y)}{P_Y(y)},
\end{align}
which is the Bayes theorem. Plugging Kronecker deltas in the place of
$a(x)$ leads to the posterior probability distribution of $X$ given
$Y = y$.

\subsection{Nondemolition principle (NDP)}
I now review Belavkin's NDP for quantum inference.  To focus on the
physics and avoid cumbersome mathematical technicalities, assume that
all the Hilbert spaces considered hereafter (until
Sec.~\ref{sec_gauss}) are finite-dimensional, so that the principle
becomes especially simple \cite{bouten}. Let $\mathcal O(\mathcal H)$
be the set of operators on a Hilbert space $\mathcal H$ and $\rho$ a
density operator on $\mathcal H$.  Let $A \in \mathcal O(\mathcal H)$
be the hidden observable to be estimated and
$B\in \mathcal O(\mathcal H)$ an operator-valued estimator, both in
the Heisenberg picture. Assume that both are Hermitian.  Typically,
$B$ is restricted to come from a subspace of $\mathcal O(\mathcal
H)$. The NDP demands that $A$ and $B$ commute, that is,
$[A,B] \equiv AB-BA = 0$.  Then there exists a common orthonormal
basis $\BK{\ket{\omega} \in \mathcal H:\omega \in \Omega}$ such that
$A$ and $B$ can be simultaneously diagonalized as
\begin{align}
A &= \sum_\omega a(\omega)\ket{\omega}\bra{\omega},
&
B &= \sum_\omega b(\omega)\ket{\omega}\bra{\omega}
\end{align}
in terms of some functions $a,b:\Omega \to \mathbb R$.  The physical
meaning of the compatibility is that $A$ and $B$ can be jointly
measured by external classical observers in the same experiment, and
the observers can compare the outcomes to evaluate the quality of the
estimator. More generally, one may use a set of commuting observables
that are measured to construct $B$.  With the common basis, all the
outcomes observe the probability measure
\begin{align}
P(\omega) &= \bra{\omega}\rho\ket{\omega}.
\end{align}
Classical probability theory, as well as the classical conditional
expectation presented in Section~\ref{sec_ce}, can then be applied to
the inference problem.

One key application of the NDP is Belavkin's approach to quantum
filtering, in which $B$ is a function of the observables of a field
probing a system and $A$ is an observable of the system ahead of the
field measurements. The compatibility among the observables in the
Heisenberg picture can be proved for a general class of Markovian
models \cite{belavkin_qnd}; see also Sec.~3.2.2 in
Ref.~\cite{gardiner_zoller}. The formula for the conditional
expectation of an arbitrary $A$ can then be used to derive the
stochastic master equation that governs the posterior quantum state
\cite{belavkin_qnd,bouten}.

While the NDP has been successful in producing a mathematically
satisfying theory of quantum filtering, it is, at its physical core,
nothing but the orthodox quantum measurement theory that goes back to
von Neumann. The use of only standard concepts is in fact a key virtue
of Belavkin's work, as it clarifies that no extension to standard
quantum mechanics is needed to solve his problem
\cite{belavkin_qnd}. On the other hand, Belavkin's writings do not
seem to express any strong opinion about what one should or should not
do with incompatible observables. Some researchers following his work
have nonetheless adopted a harder line \cite{gough20,private}, arguing
that quantum inference should be done \emph{only} if the NDP is
observed, and retrodictive questions do not make sense, such as the
question of which slit a photon goes through in a two-slit
interference experiment \cite{gough20}.

There is no doubt that the NDP agrees with standard quantum
measurement theory and quantum inference methods based on the NDP
agree with classical probability theory. It is debatable, however,
whether inference methods that violate the NDP should be strictly
forbidden. Despite the strong view of some, the physics community has
continued to demonstrate significant interest in such ``forbidden''
problems \cite{aav,kocsis11}.

\section{\label{sec_gce}Generalized conditional expectations}

\subsection{Inner products and Hilbert spaces for operators}
The mathematical physics literature has recognized other ways of
generalizing the conditional expectation in quantum mechanics
\cite{accardi82,petz86,petz88a,jencova10,hayashi}. One route is to
generalize the Hilbert-space treatment of random variables in
Sec.~\ref{sec_ce} for operators. Given two operators
$A,B \in \mathcal O(\mathcal H)$, define an inner product and a norm
as
\begin{align}
\Avg{B,A}_\rho &\equiv \trace B^\dagger \product_\rho A,
\label{inner}
\\
\norm{A}_\rho &\equiv \sqrt{\Avg{A,A}_\rho},
\end{align}
where $\dagger$ denotes the adjoint, $\trace$ denotes the trace, and
$\product_\rho:\mathcal O(\mathcal H)\to \mathcal O(\mathcal H)$ is a
linear map that depends on $\rho$. Assume that $\product_\rho$ is
self-adjoint and positive-semidefinite with respect to the
Hilbert-Schmidt inner product
\begin{align}
  \Avg{B,A}_{\mathrm{HS}} \equiv \trace B^\dagger A,
\end{align}
such that Eq.~(\ref{inner}) also qualifies as an inner product.
Assume further that $\product$ satisfies the following properties:
\begin{align}
\product_\rho A &= \rho A
\textrm{ if $\rho$ and $A$ commute,}
\label{commute}
\\
\product_\rho (U^\dagger A U) &= U^\dagger \bk{\product_{U\rho U^\dagger}  A} U,
\label{unitary}
\\
\product_{\rho\otimes \rho'}  (A\otimes A') &= (\product_\rho A) \otimes (\product_{\rho'} A'),
\label{tensor}
\end{align}
where $U$ is a unitary operator, $\rho'$ is another density operator
on Hilbert space $\mathcal H'$, and $A' \in\mathcal O(\mathcal H')$.
Equation~(\ref{commute}) ensures that the inner product coincides with
the classical version for commuting operators.
Equations~(\ref{unitary}) and (\ref{tensor}) are desirable properties
in dealing with dynamics and composite Hilbert spaces \cite{hayashi}.
In what follows, I further require that
\begin{align}
\norm{A\otimes I}_\rho &= \norm{A}_{\trace'\rho},
\label{variance}
\end{align}
where $I$ is the identity operator, $\rho$ is a density operator on
$\mathcal H\otimes \mathcal H'$, and $\trace'$ is the partial trace
over $\mathcal H'$. Equation~(\ref{variance}) is a reasonable
requirement for the definition of a quantum variance.  Given these
properties, $\product$ is not unique. Some prominent examples that
satisfy Eqs.~(\ref{commute})--(\ref{tensor}) include the left product
\begin{align}
\product_\rho A &= \rho A,
\end{align}
the Jordan product
\begin{align}
\product_\rho A &= \frac{1}{2} \bk{\rho A + A \rho},
\end{align}
and the root product
\begin{align}
\product_\rho A &= \sqrt{\rho} A \sqrt{\rho}.
\label{root}
\end{align}
More generally, a class of $\product$ can be constructed from convex
combinations of $\product_\rho A = \rho^{\lambda}A\rho^{1-\lambda}$,
$0\le\lambda\le 1$ \cite{hayashi,petz93}. With the left product,
Eq.~(\ref{inner}) becomes the inner product in the
Gelfand-Naimark-Segal construction \cite{hamhalter}.  With the Jordan
product, Eq.~(\ref{inner}) becomes an inner product proposed by Holevo
that is useful in quantum statistics \cite{holevo77,holevo11}.  The
other products also have their uses in mathematical physics and
statistical mechanics \cite{petz88a,jencova10,petz93}.  The left
product and the Jordan product further satisfy Eq.~(\ref{variance}),
although the root product and many others do not \cite{hayashi}.

With an inner product and the associated norm at hand, an operator
Hilbert space $L_2(\rho)$ can be constructed from
$\mathcal O(\mathcal H)$, generalizing the classical $L_2(P)$ space
described in Sec.~\ref{sec_ce}. Because $\mathcal H$ is
finite-dimensional, there is no need to complete the space with
unbounded operators \cite{holevo11}.

A distance between two operators can be defined as
\begin{align}
  d_\rho(A,B) &\equiv \norm{A - B}_\rho.
\label{drho}
\end{align}
In the context of quantum inference, Eq.~(\ref{drho}) can serve as a
generalization of the classical root-mean-square error given by
Eq.~(\ref{dP}) as a performance criterion \cite{dressel15,ohki15}.
The choice of an estimator to minimize Eq.~(\ref{drho}) may be called
the minimum-error principle.  Note that the principle imposes no
requirement on the compatibility between $A$ and $B$. The whole point
of inference is that $A$ is hidden and can only be inferred, so the
NDP's requirement that $A$ be jointly measurable with $B$ by external
classical observers may seem too restrictive if $A$ is never measured
in reality. The minimum-error principle, on the other hand, avoids the
stringent requirement and serves as a more general principle for
quantum inference beyond an exact correspondence with probability
theory. To quote Belavkin himself on the limitation of quantum
probability theory \cite{belavkin_qnd}:
\begin{quote}
  It is nonsense to consider seriously a complete observation in the
  closed universe; there is no universal quantum observation, no
  universal reduction and spontaneous localization for the wave
  function of the world. Nobody can prepare an a priori state
  compatible with a complete world observation and reduce the a
  posteriori state, except God. But acceptance of God as an external
  subject of the physical world is at variance with the closeness
  assumption of the universe. Thus, the world state-vector has no
  statistical interpretation, and the humanitarian validity of these
  interpretations would, in any case, be zero. The probabilistic
  interpretation of the state-vector is relevant to only the induced
  states of the quantum open objects being prepared by
  experimentalists in an appropriate compound system for the
  nondemolition observation to produce the reduced states after the
  registration.
\end{quote}
Unless we impose an artificial classical-quantum boundary or force
quantum mechanics to serve our classical intuition, nothing in quantum
mechanics mandates that observables should commute.

\subsection{Application of open quantum system theory}
To treat time evolution and open quantum systems, suppose that
\begin{align}
\mathcal H = \mathcal H_1\otimes \mathcal H_2\otimes \mathcal H_3
\end{align}
consists of three Hilbert subspaces $\mathcal H_1$, $\mathcal H_2$,
and $\mathcal H_3$, 
\begin{align}
\rho = \sigma\otimes \tau
\end{align}
is the density operator at, say, time $t$ with $\sigma$ being a
density operator on $\mathcal H_1$ and $\tau$ a density operator on
$\mathcal H_2\otimes\mathcal H_3$,
\begin{align}
A_t \equiv A\otimes I \otimes I
\label{At}
\end{align}
is the hidden observable with $A \in \mathcal O(\mathcal H_1)$,
\begin{align}
B_T \equiv U^\dagger (I\otimes B\otimes I) U
\label{BT}
\end{align}
is the Heisenberg picture of an operator-valued estimator
$B \in \mathcal O(\mathcal H_2)$ at time $T \ge t$, and $U$ is a
unitary operator on $\mathcal H$ that models the time evolution from
$t$ to $T$. The mean-square error of $B_T$ in inferring
$A_t$ becomes
\begin{align}
d_\rho^2(A_t,B_T) &= 
\norm{A}_\sigma^2 - 2 \real \Avg{\cp^\dagger B,A}_\sigma 
\nonumber\\&\quad + 
\norm{I \otimes B\otimes I}_{U\rho U^\dagger}^2,
\end{align}
where 
\begin{align}
\cp \sigma &\equiv  \trace_{13} U(\sigma\otimes\tau)U^\dagger
\label{tpcp}
\end{align}
is a trace-preserving completely positive (TPCP) map
$\cp : \mathcal O(\mathcal H_1)\to \mathcal O(\mathcal H_2)$ in the
Stinespring representation \cite{hayashi}, $\cp^\dagger$ is its
adjoint with respect to the Hilbert-Schmidt inner product,
$\trace_{13}$ denotes the partial trace over the Hilbert spaces
numbered by the subscript ($\mathcal H_1\otimes\mathcal H_3$ here),
and the properties given by Eqs.~(\ref{unitary}) and (\ref{tensor})
have been used.  If Eq.~(\ref{variance}) is also used, then
\begin{align}
d_\rho^2(A_t,B_T) &= D_{\sigma,\cp}(A,B), 
\label{drho_hei}
\\
D_{\sigma,\cp}(A,B)
&\equiv
\norm{A}_\sigma^2 - 2 \real \Avg{\cp^\dagger B,A}_\sigma 
%\nonumber\\&\quad
+ \norm{B}_{\cp \sigma}^2.
\label{mse_open}
\end{align}
Let $B = \gce A$ be an estimator of $A$ that minimizes
Eq.~(\ref{mse_open}). By substituting $B = \gce A + \epsilon c$ into
Eq.~(\ref{mse_open}), differentiating with respect to $\epsilon$, and
assuming that $c$ is an arbitrary operator, it is straightforward to
show that $\gce A$ obeys
\begin{align}
\Avg{\cp^\dagger c,A}_\sigma &= \Avg{c,\gce A}_{\cp\sigma}
\quad
\forall c \in L_2(\cp\sigma).
\label{inner_gce}
\end{align}
Notice that this equation is a generalization of
Eq.~(\ref{inner_ce3}), with $\sigma$ generalizing the probability
measure $P$, $\cp\sigma$ generalizing the coarse-grained measure
$P_Y$, $\cp^\dagger c$ generalizing the pullback $Y^* c$, and
$\gce A$ generalizing the conditional expectation $\check a$. The
existence and uniqueness of $\gce A$ as an element in $L_2(\cp\sigma)$
is guaranteed by the Riesz representation theorem
\cite{parthasarathy05}, if the left-hand side of Eq.~(\ref{inner_gce})
is regarded as a linear functional of $c$, necessarily bounded in the
finite-dimensional case considered here.

Equation~(\ref{inner_gce}) can be expressed in terms of the
Hilbert-Schmidt inner product as
\begin{align}
\Avg{\cp^\dagger c,\product_\sigma A}_{\mathrm{HS}}
= \Avg{c,\cp\product_\sigma A}_{\mathrm{HS}}
= \Avg{c,\product_{\cp\sigma} \gce A}_{\mathrm{HS}}.
\end{align}
Since $c$ is arbitrary, the equation is reduced to
\begin{align}
\product_{\cp \sigma} \gce A &= \cp \product_{\sigma} A.
\label{GCE}
\end{align}
Equation~(\ref{GCE}) coincides with the general definition of a GCE
given by Eq.~(6.21) in Ref.~\cite{hayashi}; see also
Ref.~\cite{petz88a}.  If $\product_{\cp\sigma}$ is invertible,
$\gce:\mathcal O(\mathcal H_1) \to \mathcal O(\mathcal H_2)$ as a map
is explicitly given by
\begin{align}
\gce = \product_{\cp\sigma}^{-1}\cp \product_{\sigma}.
\label{GCE2}
\end{align}
Any $\gce A$ that satisfies Eq.~(\ref{GCE}) leads to
\begin{align}
\min_{B \in L_2(\cp\sigma)}D_{\sigma,\cp}(A,B)
&= \norm{A}_\sigma^2 - \norm{\gce A}_{\cp \sigma}^2.
\label{mmse}
\end{align}
Equation~(\ref{GCE}) can be used to define a GCE even if the
$\product$ map does not satisfy Eq.~(\ref{variance}). For example, the
Accardi-Cecchini GCE \cite{accardi82} given by
\begin{align}
\cp_\sigma A
&= \bk{\cp\sigma}^{-1/2}\cp(\sqrt{\sigma}A\sqrt{\sigma})\bk{\cp\sigma}^{-1/2}
\end{align}
results from Eq.~(\ref{GCE2}) if $\product$ is the root product given
by Eq.~(\ref{root}) \cite{petz88a,jencova10}. Then the GCE still has
the meaning of an operator that minimizes the $D$ in
Eq.~(\ref{mse_open}). As long as $\product$ leads to
$\norm{I\otimes B\otimes I}_{U\rho U^\dagger} \le
\norm{B}_{\cp\sigma}$ and thus
$d_\rho^2(A_t,B_T) \le D_{\sigma,\cp}(A,B)$, Eq.~(\ref{mmse}) and
therefore Eq.~(\ref{mse_open}) remain nonnegative \cite{hayashi}.

It is straightforward to show that many versions of the $\product$
map satisfy the property of mapping Hermitian operators to Hermitian
operators. An example is
\begin{align}
\product_\rho A &= 
\frac{1}{2}\bk{\rho^\lambda A \rho^{1-\lambda} + \rho^{1-\lambda} A \rho^\lambda},
&
0 \le \lambda \le 1,
\end{align}
which includes the Jordan product and the root product. More
generally, any convex combination of $\product$'s satisfying the
property also satisfies it. The inner product then becomes real if the
operators are restricted to be Hermitian. Since any completely
positive map $\cp$ also satisfies such a property (as easily proved
using its Kraus representation), the GCE given by Eq.~(\ref{GCE2})
associated with such an $\product$ map also maps Hermitian operators
to Hermitian operators, a property that some may find desirable.

At this juncture, the works by Leifer and Spekkens \cite{leifer13} and
Horsman and coworkers \cite{horsman17} on two-time quantum states
deserve a mention. They diverged from the formalism here by expressing
the TPCP map in its Choi form, although they also recognized the
usefulness of the root product or the Jordan product in their
formalism. It is outside the scope of this paper to investigate the
connections of these works to the ones considered here.

\subsection{Discussion}
Among the choices of $\product$ considered here, the Jordan product
seems to stand out as the most reasonable, as it satisfies all the
desirable properties given by Eqs.~(\ref{commute})--(\ref{variance})
and gives a Hermitian GCE for a Hermitian hidden observable. With the
Jordan product and Hermitian $A_t$ and $B_T$,
the error given by Eq.~(\ref{drho}) becomes
\begin{align}
d_\rho^2(A_t,B_T) &= \trace \rho (A_t-B_T)^2.
\end{align}
$A_t-B_T$ is another Hermitian observable that can be measured in
principle, and the resulting variance coincides with the error, so the
error does have a probabilistic interpretation. Although a measurement
of $A_t-B_T$ seems difficult in practice, this interpretation does not
appear to be much less reasonable than the interpretation of the NDP
in terms of the joint measurement of $A_t$ and $B_T$, which is also
quite impractical.

Along the same line, for parameter estimation problems, where $A$
models a classical real parameter, the Jordan product has the
desirable feature of making the error agree with the classical
estimation error upon the measurement of a Hermitian $B$
\cite{personick71,belavkin73}.

In the more mathematical context of quantum information theory, the
Jordan product leads to the smallest---and thus the most
useful---quantum version of the Fisher information for scalar
parameter estimation \cite{petz96,hayashi}. The GCE given by
Eq.~(\ref{GCE}) determines the relation between the scores
(logarithmic derivative operators) of a quantum parametric model
before and after the $\cp$ map, generalizing the classical case (see,
for example, Sec.~25.5 in Ref.~\cite{vaart}), and the fact that $\gce$
is a projection can be used to prove the monotonicity of the quantum
Fisher information \cite{hayashi,tsang21a}.

Other problems may require a different product or a different GCE,
however.  For example, the left product leads to a quantum Fisher
information matrix that may be more useful for vectoral parameter
estimation \cite{holevo11,yuen_lax}, while the Accardi-Cecchini GCE
and its Hilbert-Schmidt adjoint, the Petz recovery map, are central to
the study of quantum channel sufficiency
\cite{petz86,petz88a,jencova10}. There is no single best definition of
a GCE, nor does there need to be, just as there is no single best
quasiprobability representation for every problem.

\section{\label{sec_hybrid}Examples that obey the nondemolition
  principle}
\subsection{\label{sec_classical}Classical}
The classical case is unequivocal. Let $P_X(x)$ be the prior
probability distribution of a hidden random variable
$X \in \mathcal X$ and $P_{Y|X}(y|x)$ be the probability distribution
of the observed $Y \in \mathcal Y$ conditioned on $X = x$. For
simplicity, assume that $\mathcal X$ and $\mathcal Y$ are finite sets
and all the probabilities are positive.  Then
$P_Y(y) = \sum_x P_{Y|X}(y|x)P_X(x) > 0$. Let
\begin{align}
\sigma &= \sum_x P_X(x)\ket{x}\bra{x},
\label{classical_sigma}
\\
A &= \sum_x a(x)\ket{x}\bra{x},
\label{classical_A}
\\
\cp \sigma &= \sum_{y,x} P_{Y|X}(y|x) \bra{x}\sigma\ket{x} \ket{y}\bra{y},
\\
\gce A &= \sum_y \check a(y)\ket{y}\bra{y},
\label{classical_est}
\end{align}
where $\{\ket{x}:x \in \mathcal X\}$ is an orthonormal basis of
$\mathcal H_1$, $\{\ket{y}: y \in \mathcal Y\}$ is an orthonormal
basis of $\mathcal H_2$, $a:\mathcal X \to \mathbb C$ is a classical
random variable, and $\check a:\mathcal Y \to \mathbb C$ is an
estimator given the observation.  The GCE given by Eq.~(\ref{GCE2}),
regardless of the choice of the $\product$ map, becomes
\begin{align}
\check a(y) &= \sum_{x} \frac{P_{Y|X}(y|x) P_X(x)}{P_Y(y)} a(x),
\label{bayes}
\end{align}
which is, of course, the Bayes theorem.

\subsection{\label{sec_retro}Retrodiction}
For an example with quantum ingredients, consider the retrodiction
problem studied by Watanabe \cite{watanabe} and Barnett, Pegg, and
Jeffers \cite{barnett21}. Assume the classical model given by
Eqs.~(\ref{classical_sigma})--(\ref{bayes}). In addition, assume that
the observation distribution $P_{Y|X}(y|x)$ arises from Born's rule,
viz.,
\begin{align}
P_{Y|X}(y|x) &= \trace M(y) \mathcal G \rho_x 
= \Avg{M(y),\mathcal G \rho_x}_{\mathrm{HS}}
\label{retro}
\\
&= \Avg{\mathcal G^\dagger M(y),\rho_x}_{\mathrm{HS}},
\end{align}
where $\rho_x$ is the density operator of a quantum system conditioned
on $X = x$ at an initial time, $M$ is the positive operator-valued
measure (POVM) that models a measurement of the system at a final
time, and $\mathcal G$ is a TPCP map that models the time evolution
in-between. As recognized by Refs.~\cite{watanabe,barnett21}, one can
choose to evolve $\rho_x$ forward in time by $\mathcal G$ or evolve
$M(y)$ backward in time by $\mathcal G^\dagger$.  Either way, the GCE
for this retrodiction problem is still the classical Bayes theorem
given by Eq.~(\ref{bayes}) and can be written as
\begin{align}
\check a(y)  &= 
\sum_x \frac{\trace M(y)\mathcal G \rho(x)}
{\sum_{x'} \trace M(y)\mathcal G \rho(x')} a(x),
\\
\rho(x) &\equiv \rho_x P_X(x),
\end{align}
where $\rho(x)$ is the hybrid density operator \cite{wiseman_milburn}.
In this context, there is some freedom in how one normalizes $M(y)$
and $\rho(x)$ \cite{barnett21}.

Although Barnett and coworkers called this problem quantum
retrodiction, here it may be more appropriate to specify it as hybrid
retrodiction, since the hidden observable is classical, while the
observation arises from Born's rule in quantum mechanics.

\subsection{\label{sec_smooth}Hybrid smoothing}
A generalization of the retrodiction problem is the so-called
smoothing problem \cite{kalman,sarkka}, where the value of a
time-varying waveform at a certain intermediate time $t$ is to be
estimated using observations both before and after $t$.  Consider in
particular a classical waveform, such as a gravitational wave,
perturbing a quantum system, such as an optomechanical sensor.
Sequential or continuous measurements are made on the quantum system,
and the outcomes are used to infer the waveform. To model this hybrid
smoothing problem, assume again the classical model given by
Eqs.~(\ref{classical_sigma})--(\ref{bayes}). Let the classical
waveform value at an intermediate time $t$ be $X$ and the measurement
model be
\begin{align}
P_{Y|X}(y|x) &= \trace M(y|x) \rho_x,
\end{align}
where $\rho_x$ is now the density operator of the quantum system at
time $t$ conditioned on $X = x$, the POVM $M(y|x)$ is also conditioned
on $X = x$, and $Y$ represents all the ``future'' observations after
$t$. All quantities, including $P_X(x)$, are implicitly assumed to be
conditioned on the ``past'' observations before $t$.  The GCE becomes
\begin{align}
\check a(y)  &= 
\sum_x \frac{\trace M(y|x)\rho(x)}{\sum_{x'} \trace M(y|x')\rho(x')} a(x),
\label{hybrid_smooth}
\end{align}
which is the central formula employed by Tsang in his hybrid smoothing
theory \cite{smooth,smooth_pra1} (with the obvious generalization of
$P_X(x)$ to a density and $\sum_x$ to an integral).

Hybrid filtering refers to the estimation of $X$ with observations up
to time $t$ only and can be accomplished by computing the $\rho(x)$
conditioned on the past observations \cite{wiseman_milburn}. As is
well known in engineering \cite{sarkka}, smoothing is more accurate
than filtering if the waveform is stochastic, since the future
observations contain information about the waveform value that is
absent in the past observations.  These considerations arguably make
the hybrid smoothing theory the most useful offshoot of the quantum
retrodiction and smoothing formalism.

If a continuous measurement is performed on the quantum system,
$M(y|x)$ and $\rho(x)$ can be solved via a time-symmetric pair of
stochastic master equations, as proposed by Tsang
\cite{smooth,smooth_pra1}. $\rho(x)$ is to be solved using a
forward-time stochastic master equation for hybrid filtering that goes
from an initial time to the intermediate time $t$, while $M(y|x)$ is
to be solved using an adjoint equation that goes backward from a final
time to $t$, generalizing the retrodiction formalism in
Sec.~\ref{sec_retro}.

Equation~(\ref{hybrid_smooth}) may be solved more efficiently by
considering the quasiprobability representations of $M(y|x)$ and
$\rho(x)$, which can admit more succinct forms in special cases.  For
example, for linear Guassian systems, the Wigner representations of
$M(y|x)$ and $\rho(x)$ are both Gaussian, and the smoother coincides
with the optimal linear smoother in the classical setting
\cite{smooth,smooth_pra1,sarkka}.  The ``Gaussian theory of
hindsight'' proposed earlier by Petersen and M{\o}lmer for atomic
magnetometry \cite{petersen} can then be viewed as a special case.
Tsang, Wiseman, and Caves showed that the smoothing technique is
needed to achieve the fundamental quantum limit to waveform estimation
in optomechanical force sensing \cite{twc}.
References~\cite{wheatley,*yonezawa,*iwasawa,*wheatley15} report
experimental demonstrations of the smoothing technique for quantum
optical systems.

The more recent proposal of ``past quantum state'' by Gammelmark,
Julsgaard, and M{\o}lmer \cite{gammelmark2013} is nothing but a
special case of the hybrid smoothing theory. Let
$\{\kappa_{x}:x \in \mathcal X\}$ be a set of Kraus operators that
model a measurement at time $t$ and $\rho'$ be the quantum state
before the measurement. Their theory can be reproduced from
Eq.~(\ref{hybrid_smooth}) by assuming
\begin{align}
\rho_x &= \frac{\kappa_{x} \rho' \kappa_{x}^\dagger}
{\trace \kappa_{x} \rho' \kappa_{x}^\dagger},
\quad
P_X(x) = \trace \kappa_{x} \rho' \kappa_{x}^\dagger,
\\
\rho(x) &= \rho_x P_X(x) = \kappa_{x} \rho' \kappa_{x}^\dagger.
\end{align}
In other words, the measurement outcome $X$ that is assumed to be
hidden in their setup can simply be treated as a classical random
variable in the hybrid theory; see also Ref.~\cite{gough20}. M{\o}lmer
and coworkers have since produced a series of papers on the subject
\cite{chantasri21}, such as Ref.~\cite{madsen21}, which rediscovers
the hybrid smoothing theory.

The quantum state smoothing theory proposed by Guevara and Wiseman
\cite{guevara,chantasri21}, which concerns the estimation of a
time-dependent density matrix with partial observations, may also be
considered as a special case of hybrid smoothing. In their scenario, a
partial observer named Alice has only partial access to a sequence of
observations of a quantum system, and her goal is to estimate the
density matrix possessed by an omniscient observer named Bob, who has
complete access to the observations and updates his density matrix
continuously with them.  The key is to view the problem as a
generalization of quantum state tomography, where the density matrix
is a \emph{classical} matrix-valued parameter \cite{paris04}. Under
this view, Bob's density matrix is a classical stochastic process,
whose equation of motion happens to be the stochastic master equation
driven by the observations as an effective system noise. The hybrid
approach should therefore be applicable to the state estimation
problem, although the technical details remain to be worked out.  It
is an interesting open question whether the use of other cost
functions in Ref.~\cite{chantasri21} may be applied to the hybrid
theory.

\subsection{\label{sec_personick}Optimal quantum estimation of a
  classical parameter}
A generalization of the previous examples is to allow the experimenter
to pick any quantum measurement that minimizes the error.  Keep the
classical $\sigma$ and $A$ given by Eqs.~(\ref{classical_sigma}) and
(\ref{classical_A}), but assume that the output state
\begin{align}
\cp \sigma &= \sum_x \rho_x \bra{x}\sigma\ket{x}
\end{align}
is quantum. Then the GCE given by Eq.~(\ref{GCE}) obeys
\begin{align}
\product_{\sum_x \rho(x)}\gce A &= \sum_x \rho(x) a(x).
\end{align}
If $a$ is real and $\product$ is the Jordan product, any solution for
$\gce A$ is an optimal observable to be measured for estimating
$a(x)$, as discovered by Personick \cite{personick71}; see
Ref.~\cite{macieszczak14} for a more recent related work.

\subsection{\label{sec_belavkin}Optimal quantum estimation}
The quantum estimation problem considered by Belavkin and Grishanin
\cite{belavkin73} is a further variation of the previous examples.
Let the larger Hilbert space be $\mathcal H_1\otimes\mathcal H_2$ and
the state on this Hilbert space be $\sigma$.  The parameter to be
estimated is now a quantum observable $A$ on $\mathcal H_1$, while the
estimator is another quantum observable $B$ on $\mathcal H_2$. Assume
that both are Hermitian. Their TPCP map is simply the partial trace
$\cp \sigma = \trace_2\sigma$.  The resulting optimal observable
obtained by Belavkin and Grishanin coincides with the GCE here in
terms of the Jordan product.

\subsection{\label{sec_ndp}Compliance with the nondemolition
  principle}
All the previous examples obey the NDP.  To check this explicitly for
the examples in Secs.~\ref{sec_classical}--\ref{sec_personick}, notice
that the TPCP maps there can all be expressed in the Stinespring
representation given by Eq.~(\ref{tpcp}) if one assumes
\begin{align}
\tau &=\ket{\phi}\bra{\phi},
\\
U &= \sum_x \ket{x}\bra{x}\otimes V_x,
\label{U_hybrid}
\end{align}
where $\ket{\phi}$ is a pure state in
$\mathcal H_2\otimes\mathcal H_3$, $\{\ket{x}:x \in \mathcal X\}$ is
the orthonormal basis of $\mathcal H_1$ assumed in
Secs.~\ref{sec_classical}--\ref{sec_personick}, and
$V_x \in \mathcal O(\mathcal H_2\otimes\mathcal H_3)$ is a unitary
operator controlled by $x$. The TPCP map becomes
\begin{align}
\cp\sigma &= 
\sum_x \bra{x}\sigma\ket{x} \trace_3 V_x\ket{\phi}\bra{\phi} V_x^\dagger.
\end{align}
The classical model in Sec.~\ref{sec_classical}, the retrodiction
model in Sec.~\ref{sec_retro}, and the hybrid smoothing model in
Sec.~\ref{sec_smooth} are obtained if
\begin{align}
\trace_3 V_x\ket{\phi}\bra{\phi} V_x^\dagger &=
\sum_y P_{Y|X}(y|x)\ket{y}\bra{y},
\label{purification1}
\end{align}
while Personick's model in Sec.~\ref{sec_personick} is obtained if
\begin{align}
\trace_3 V_x\ket{\phi}\bra{\phi} V_x^\dagger &= \rho_x.
\label{purification2}
\end{align}
Given the density operator on the right-hand side of
Eq.~(\ref{purification1}) or Eq.~(\ref{purification2}), standard open
quantum system theory \cite{hayashi} assures that there always exist a
purification of the density operator and a unitary $V_x$ that maps
$\ket{\phi}$ to the purification. The fact that $P_{Y|X}(y|x)$ in the
hybrid models arises from Born's rule turns out to be irrelevant to
the compatibility between the hidden observable and the estimator.

In the larger Hilbert space
$\mathcal H_1\otimes\mathcal H_2\otimes\mathcal H_3$, the hidden
observable $A_t$ and the estimator observable $B_T$ in the Heisenberg
picture are given by Eqs.~(\ref{At}) and (\ref{BT}) respectively.  The
classical $A$ given by Eq.~(\ref{classical_A}) becomes
\begin{align}
A_t &= \sum_x a(x)\ket{x}\bra{x}\otimes I\otimes I,
\end{align}
while applying the $U$ given by Eq.~(\ref{U_hybrid})
to Eq.~(\ref{BT}) leads to
\begin{align}
B_T &= \sum_x\ket{x}\bra{x} \otimes V_x^\dagger(B\otimes I)V_x.
\end{align}
The $A_t$ and $B_T$ here commute, meaning that the NDP is observed.

To treat infinite-dimensional problems, the obvious generalization
would be to replace $\ket{x}\bra{x}$ by a projection-valued measure
and $\sum_x$ by an integral, although a rigorous analysis of the
infinite-dimensional case is outside the scope of this paper.

For the problem considered by Belavkin and Grishanin and described in
Sec.~\ref{sec_belavkin}, the observables in the larger Hilbert space
are simply $A\otimes I$ and $I\otimes B$, so their compatibility is
obvious.

\section{\label{sec_nonclassical}Examples that may violate the nondemolition principle}
\subsection{\label{sec_weak}Weak values}
For an example that may violate the NDP, let $\sigma$ and $A$ be
quantum, $A$ be Hermitian, $\cp$ be the measurement map given by
\begin{align}
\cp \sigma &=\sum_y \Bk{\trace M(y)\sigma} \ket{y}\bra{y},
\label{measure_map}
\end{align}
and the estimator be the classical form given by
Eq.~(\ref{classical_est}). Then Eq.~(\ref{GCE2}) becomes
\begin{align}
\check a(y) &= \frac{\trace M(y)\product_\sigma A}{\trace M(y)\sigma}.
\label{weak}
\end{align}
This GCE coincides with the weak value if $\product$ is the left
product and the real part of the weak value if $\product$ is the
Jordan product \cite{ohki15}.  The weak values may be generalized by
considering other operator products for $\product$, while a larger
class of quasiprobability distributions that are conditioned on both
past and future observations may be generated by plugging quantum
generalizations of the Kronecker deltas in the place of $A$, such as
projectors \cite{dressel15} or phase-point operators \cite{ferrie11}.

As the weak value is well known to create paradoxes that defy
classical logic \cite{aav}, one should not expect it to obey the NDP
in general. To check, notice that the measurement map given by
Eq.~(\ref{measure_map}) can be expressed in the Stinespring
representation given by Eq.~(\ref{tpcp}) if one assumes
\begin{align}
M(y) &= \trace_{3} \Pi_{13}(y) (I \otimes \tau_3),
\label{naimark}
  \\
  \tau &= \ket{\phi}\bra{\phi}\otimes \tau_3,
  \\
  U &= \sum_y \Pi_{13}(y)\otimes V_y,
\label{U_measure}
\end{align}
where $\Pi_{13}\in\mathcal O(\mathcal H_1\otimes\mathcal H_3)$ is the
projection-valued measure and $\tau_3 \in \mathcal O(\mathcal H_3)$ is
the ancilla state that arise from the Naimark extension of $M(y)$
\cite{hayashi}, while $V_y \in \mathcal O(\mathcal H_2)$ is a unitary
that gives $V_y\ket{\phi} = \ket{y}$. Note that Eq.~(\ref{U_measure})
is expressed in the order of
$\mathcal O(\mathcal H_1\otimes\mathcal H_3) \otimes \mathcal
O(\mathcal H_2)$. Following the same order, the operator-valued
estimator given by Eq.~(\ref{BT}) can be expressed as
\begin{align}
B_T &= \sum_y \Pi_{13}(y) \otimes V_y^\dagger B V_y,
\label{BT_weak}
\end{align}
which may not commute with the hidden observable given by
Eq.~(\ref{At}).

Despite the possible violation of the NDP, the weak values are not
necessarily paradoxical.  The literature on the weak values tends to
focus on the paradoxes, but it is arguably more important to
demonstrate that they make sense for large classes of problems, so
that they can serve as a discerning test of nonclassicality. The next
section demonstrates that, for a certain class of systems called
linear Gaussian systems, a classical probability model can indeed be
constructed even for incompatible observables, thus guaranteeing the
inference to conform with classical logic.

\subsection{\label{sec_gauss}Linear Gaussian systems}
A large class of quantum optics experiments, such as optomechanics,
can be modeled as linear Gaussian systems
\cite{wiseman_milburn,holevo19,weedbrook}.  They are defined by the
following conditions:
\begin{enumerate}
\item The Wigner representations of all the density operators involved
  are Gaussian.
\item The observables of interest are restricted to quadrature
  operators, defined as real linear combinations of canonical position
  and momentum operators.

\item All the unitary operators involved are generated by Hamiltonians
  that are quadratic with respect to the quadrature operators, such
  that the equations of motion for the quadratures are linear.

\item The measurements are restricted to spectral resolutions of the
  quadrature operators, such as homodyne detection in optics.

\end{enumerate}
These conditions can also be applied to TPCP maps and POVMs via their
Stinespring or Naimark representations.  It is well established that,
for linear Gaussian systems, the Wigner representation offers a
classical probability model for the quadratures, despite the
incompatibility among them \cite{wiseman_milburn,holevo19,weedbrook}.

Assume a quantum system with $N$ bosonic modes with density operator
$\sigma$ at time $t$. Let $Q$ be a column vector of the $2N$
phase-space (position and momentum) operators. Some of the phase-space
operators can be used to model classical continuous variables in a
hybrid problem \cite{gough,*qmfs}. Suppose that the measurement
outcomes from the system are used to estimate a quadrature operator
$A$, which can be written as
\begin{align}
A = b^\top Q,
\end{align}
where $b$ is a column vector of $2N$ constants and $\top$
denotes the transpose. Let $Y \in \mathcal Y = \mathbb R^L$ be the
noisy observation of $L$ quadratures after $t$ and $M$ be the
corresponding POVM. Assume that $M$ can be expressed as
\begin{align}
M(S) &= \int_S d^Ly \gamma(y),
\label{density}
\end{align}
where $\gamma(y)$ is an operator-valued density of $M$.  The probability density of $Y$ is then
\begin{align}
f_Y(y) = \trace \gamma(y)\sigma.
\end{align}
For example, if ideal homodyne detection is performed,
$\gamma(y) = \ket{y}\bra{y}$ in terms of the Dirac eigenkets
$\{\ket{y}: y \in \mathbb R^L,\braket{y|y'} = \delta^L(y-y')\}$ for
the $L$ quadratures \cite{weedbrook}. Of course, $\gamma(y)$ can also
incorporate the effects of more general dynamics and measurements
after $t$.

Assume the Jordan product for the $\product$ map hereafter. The
$L_2(\sigma)$ space remains well defined for this infinite-dimensional
problem if one completes the space with limit points of the space of
bounded operators \cite{holevo11}, although I ignore the mathematical
complications that may arise when generalizing the earlier
finite-dimensional results in the following, as per standard practice
in quantum optics \cite{weedbrook}.  The appropriate generalization of
the $L_2(\cp\sigma)$ space in this case is the classical $L_2(f_Y)$
space with the inner product
$\avg{b,a}_{f_Y} \equiv \int d^Ly b(y) a(y) f_Y(y)$, while the
appropriate generalization of Eq.~(\ref{weak}) is
\begin{align}
\check a(y) &= \frac{\trace \gamma(y)\product_\sigma A}{\trace \gamma(y)\sigma}.
\label{weak2}
\end{align}

Let $W_\sigma(q)$ and $W_\gamma(y|q)$ be the Wigner representations of
$\sigma$ and $\gamma(y)$, respectively, where $q$ is a column vector
of the $2N$ phase-space coordinates. For the smoothing problem,
$\sigma$ and $W_\sigma$ are implicitly assumed to be conditioned on
the past observations before $t$. It is straightforward to show that
the GCE given by Eq.~(\ref{weak2}) becomes \cite{smooth_pra1}
\begin{align}
\check a(y) &= \int  d^{2N}q W(q|y) b^\top q,
\label{GCE_quasi}
\\
W(q|y) &= \frac{W_\gamma(y|q) W_\sigma(q)}{\int d^{2N}q'W_\gamma(y|q') W_\sigma(q')},
\label{GCE_quasi2}
\\
d^{2N}q &\equiv dq_1\dots dq_{2N}.
\end{align}
Equation~(\ref{GCE_quasi2}) has the form of the Bayes theorem, with
$W_\sigma$ and $W_\gamma$ playing the roles of $P_X$ and $P_{Y|X}$ in
the classical model in Sec.~\ref{sec_classical}.

For a linear Gaussian system, the Gaussianity of $W_\sigma$ and
$W_\gamma$ can be proved \cite{holevo19,weedbrook}. Assume
\begin{align}
  W_\sigma(q) &\propto 
\exp\Bk{-\frac{1}{2} (q - \check q_\sigma)^\top K_\sigma^{-1}(q-\check q_\sigma)},
\\
W_\gamma(y|q) &\propto 
\exp\Bk{-\frac{1}{2} \bk{y-h q}^\top R^{-1}\bk{y-h q}},
\end{align}
where $\check q_\sigma$ and $K_\sigma$ are the mean vector and
covariance matrix of $W_\sigma$, $y$ is an $L$-dimensional column
vector, $h$ is an $L\times 2N$ matrix, and $R$ is an $L\times L$
covariance matrix. Since both $W_\sigma$ and $W_\gamma$ are positive
here, the GCE has all the nice properties of a classical conditional
expectation, even if the NDP is violated.

Assume that $h^\top R^{-1}h$ is positive-definite, such that its
inverse is defined. Let
\begin{align}
 W_\gamma(y|q) &\propto 
\exp\Bk{-\frac{1}{2} \bk{q - \check q_\gamma}^\top K_\gamma^{-1}\bk{q-\check q_\gamma}},
\\
\check q_\gamma &= K_\gamma h^\top R^{-1} y,
\\
K_\gamma &= \bk{h^\top R^{-1}h}^{-1}.
\end{align}
Then the posterior distribution given by Eq.~(\ref{GCE_quasi2})
becomes
\begin{align}
W(q|y) &\propto 
\exp\Bk{-\frac{1}{2} (q - \check q)^\top K^{-1}(q-\check q)},
\\
\check q &= K
\bk{K_\sigma^{-1}\check q_\sigma + K_\gamma^{-1}\check q_\gamma},
\\
K &= \bk{K_\sigma^{-1}+K_\gamma^{-1}}^{-1},
\label{K_smooth}
\end{align}
and the GCE of $A$ becomes
\begin{align}
\check a &= b^\top \check q.
\end{align}
The mean-square error is given by Eqs.~(\ref{drho_hei}) and
(\ref{mmse}), and one can use the correspondence between the operator
inner product $\avg{\cdot,\cdot}_\sigma$ for quadratures and the
classical inner product with respect to the Wigner function
$\avg{\cdot,\cdot}_{W_\sigma}$ \cite{wiseman_milburn} to obtain
\begin{align}
d_\rho^2 &= b^\top K b.
\end{align}
Note that $\check q$ is the vectoral GCE of the quadrature operator
vector $Q$ and may violate the NDP as an estimator, since it is a
function of the retrodictive estimator $\check q_\gamma$ and thus the
future observations.  Despite the controversial status of $\check q$,
the hidden observable of interest $A$ may still obey the NDP with
respect to the measurement, in a hybrid smoothing problem for
example. Then no one can deny the usefulness of the retrodictive
estimator, as an intermediate tool at least, in determining the final
conditional expectation of $A$.  Compared with the filtering estimator
and its error based on $\sigma$ alone, given by
\begin{align}
\check a &= b^\top \check q_\sigma,
&
d_\rho^2 &= b^\top K_\sigma b,
\end{align}
the advantage of smoothing may be substantial.

As noticed first by Refs.~\cite{tsl2009,smooth_pra1}, the posterior
covariance matrix $K$ given by Eq.~(\ref{K_smooth}), unlike $K_\sigma$
or $K_\gamma$, may violate the Heisenberg uncertainty relation, but it
is not a paradox from the perspective of the effective classical model
with the positive Wigner functions.

\section{\label{sec_conclu}Conclusion}
I have presented a unifying theory of generalized conditional
expectations (GCEs) for quantum retrodiction and smoothing.  It is
fair to say that I have drawn on many prior works in establishing the
theory and constructing the examples.  Rather than proposing yet
another approach to the problem, the key contribution of this paper is
to make hitherto unappreciated connections among the existing works.
As these works come from diverse fields in physics and engineering and
have different motivations, it is worth pointing out that they share a
common thread and offer complementary perspectives on the quantum
retrodiction and smoothing problem.

On one hand, the GCEs are shown to be natural consequences of
generalizing the Hilbert-space treatment of the classical conditional
expectation. On the other, they are shown to coincide with many
quantum estimation methods that have philosophical as well as
engineering implications. These considerations establish the GCEs as a
principled and sensible concept for quantum inference, enabling one to
address questions previously forbidden by the nondemolition principle
(NDP). Providing answers to such questions, even if they seem
paradoxical in special cases, may serve as a mental aid for
researchers and students to develop insights and intuition about
quantum systems.  On a more practical level, the hybrid smoother,
discussed in Secs.~\ref{sec_smooth} and \ref{sec_gauss}, illustrates
how an inference about the past of a quantum sensor can improve the
estimation of a classical waveform.  Regardless of one's position on
the NDP, there is no denying that the GCEs can, at the very least,
serve as useful intermediate tools in quantum sensing and
information-processing applications.

Given the fundamental importance of the conditional expectation in
classical probability and statistics, there should be plenty of room
for the concept of GCEs to grow even further in the quantum arena.

\section*{Acknowledgment}
This research is supported by the National Research Foundation (NRF)
Singapore, under its Quantum Engineering Programme (Grant No.~QEP-P7).

\appendix

\bibliography{research2}

\end{document}